\documentstyle[12pt,epsf]{article}

\advance \topmargin by -\headheight
\advance \topmargin by -\headsep

\evensidemargin \oddsidemargin
\advance\hoffset by -3mm  % A4 is narrower.
\advance\voffset by  8mm  % A4 is taller.

\def\half{{{1\over 2}}} 
\def\ie{{\em i.e.}}

\def\np{Nucl. Phys.}

\def\prl{Phys. Rev. Lett.}
\def\pr{Phys. Rev.}
\def\ap{Ann. Phys.}

\def\ijmp{Int. J. Mod. Phys.}

\newcommand{\beq}{\begin{equation}}
\newcommand{\eeq}{\end{equation}}
\newcommand{\bear}{\begin{eqnarray}}
\newcommand{\eear}{\end{eqnarray}}

\newcommand{\MM}{{\sf M}}
\newcommand{\LL}{{\cal L}}
\newcommand{\HH}{{\cal H}}
\newcommand{\hi}{{\sf H}}
\newcommand{\dir}{{\rm D}}
\newcommand{\hphi}{{\phi^{\dag}}}
\newcommand{\HPhi}{{\Phi^{\dag}}}
\newcommand{\HPi}{{\Pi^{\dag}}}
\newcommand{\hpi}{{\pi^{\dag}}}
\newcommand{\heta}{{\eta^{\dag}}}
\newcommand{\mani}{{\cal M}}
\newcommand{\I}{{\cal I}}
\newcommand{\hc}{{\rm h.c.}}
\newcommand{\tr}{{\rm Tr}}
\newfont{\namefont}{cmr10}
\newfont{\addfont}{cmti7 scaled 1440}
\newfont{\headfontb}{cmbx10 scaled 1728}
\begin{document}
\begin{titlepage}
\begin{center} {\headfontb Hamiltonian Reduction of Non-Linear Waves}
\end{center}
\vskip 0.3truein
\begin{center}
M. Alvarez 
\end{center}
\vskip 0.2truein
\begin{center} {\addfont{Department of Physics,}}\\
{\addfont{University of Wales Swansea}}\\ {\addfont{Singleton Park, Swansea 
SA2 8PP, U.K.}}\\{\tt {e-mail: pyma@swansea.ac.uk}}
\end{center}
\vskip 1truein
\begin{center}
\bf ABSTRACT
\end{center} 
The Faddeev-Jackiw Hamiltonian Reduction approach to constrained dynamics
is applied to the collective coordinates analysis of non-linear waves, and 
compared with the alternative procedure known as symplectic formalism. 

\vskip3.5truecm
\leftline{SWAT-147  \hfill January 1997}
\leftline{hep-th/9702040}
\vskip 1in
%\begin{center}
%{\bf \Huge DRAFT}
%\end{center}
%\begin{center}
%{\bf \Large Not for Publication}
%\end{center}
\smallskip
\end{titlepage}
\setcounter{footnote}{0}

%%%%%%%%%%%%%%%%%%%%%%%%%%%%%%%%%%
%%%%%                         M A I N   T E X T                            %%%%% 
%%%%%%%%%%%%%%%%%%%%%%%%%%%%%%%%%%

\section{Introduction}
The analysis of the quantum significance of finite-energy solutions
of non-linear classical field theories has deserved much attention in the 
last two decades \cite{gj,cl,gervais,gervais2}. These solutions, which we 
shall call ``non-linear waves'' henceforth, cannot be obtained by starting
from solutions of the linear part of the field equations and treating the
non-linear terms perturbatively. For that reason, non-linear waves probe the 
non-perturbative regime of the quantum theory and are essential to our 
understanding of quantum dynamics beyond small perturbations of the vacuum. 
When non-linear waves (NLW) exist the spectrum of the quantum theory 
is divided into sectors, corresponding to excitations around the vacuum or 
around the NLWs \cite{gj}. Besides, the NLWs themselves acquire a 
particle-like status and are stabilized owing to the existence of topological 
conservation laws; these laws are associated to the boundary conditions 
imposed on the NLWs. 
 
We shall consider a general boson field theory with fields $\Phi_p$ , where $p=1,\ldots,D$. Use will be 
made of a collective coordinates decomposition of the field $\Phi$ \cite{cl, gervais,gervais2}. This
procedure is the first step in some quantization methods, but we shall not
present a full quantization of NLWs. What will be presented here is a general 
analysis of the phase space $\Gamma$ of small fluctuations around a NLW. The object of interest is 
the symplectic two-form $\Omega$ that determines the Lie algebra structure of
the phase space. As in most mechanical systems, $\Gamma$ is a cotangent bundle and the symplectic 
form is exact, $\Omega=d\,\omega$. The 1-form $\omega$ is called ``canonical 1-form'' or
``symplectic potential''. The output of this analysis is the Poisson brackets
of a set of local coordinates in $\Gamma$. 

Of course, much is already known about these matters. In most practically 
oriented approaches \cite{cl} the fluctuations about a NLW $\phi$ are restricted from the beginning 
to be orthogonal in field space to the zero modes of $\phi$. These zero modes span the tangent space of 
$\mani$, the moduli space of NLWs, at a certain point. This restriction on the fluctuations (denoted 
by $\eta$) about the classical configuration $\phi$ is solved by means of a mode decomposition of 
$\eta$. The normal modes of $\eta$ are then taken as fundamental variables. Other approaches avoid 
the mode decomposition by taking the orthogonality of $\eta$ as a constraint and applying Dirac's 
analysis of constrained dynamics \cite{tomboulis, tw}. 

Our treatment will make use of Hamiltonian Reduction. This method was put 
forward by Faddeev and Jackiw \cite{fj,jackiw} as an alternative to Dirac's 
analysis of constrained dynamics. As Dirac's, this method is concerned with 
the classical phase space of field theories and does not address the problem of
quantization, especially the issue of ordering non-commuting operators. In spite of this it is customary 
to refer to these classical analyses as ``quantization procedures''. An interesting variation on the 
Faddeev-Jackiw method, known as ``symplectic quantization'' is due to 
Wotzasek, Montani and Barcelos-Neto \cite{wotzasek,montani, barcelos}. By now 
it is clear that these new methods to quantize classical systems have
superseded Dirac's, being both simpler and more fundamental. It is thus 
interesting to see how they work in NLW quantization. There are differences
between the original Faddeev-Jackiw method and symplectic quantization, which
will be illustrated in the main text. Although our emphasis is on the use of these new methods, 
some repetition of old results is unavoidable. 

The main difference between the approach of \cite{fj, jackiw} and \cite{wotzasek, montani, barcelos}
is, roughly, that in the former we are asked to solve the constraints and reduce the phase space of the 
system to the independent degrees of freedom, while in the latter constraints are no solved but 
embedded in an extended phase space, in such a way that the constraints are non-dynamical. The 
alternative between these two versions of Hamiltonian 
Reduction is reminiscent of the alternative between imposing the constraints before of after 
quantization. In the first case we must solve the constraints and quantize the independent degrees of 
freedom, while in the second case constraints are ignored at first, but eventually we require physical 
states to be annihilated by them.  

This article is organized as follows. A very general classical theory of scalar fields 
is presented in section 2, along with its hamiltonian formulation. In section 
3 we review the collective coordinates formalism; the total field $\Phi$ is 
decomposed into a classical part $\phi$ that should be a NLW (sometimes called
the ``barion''), and a quantum part $\eta$ (the ``meson''). The classical part 
will depend on some parameters $\alpha^a$, $a=1,\ldots,N$ that are the 
collective coordinates. The dynamical variables will be $\eta$ and the collective coordinates of the 
NLW. The resulting dynamical system will be shown to be constrained. Section 4 deals 
with the Hamiltonian Reduction of this constrained system in the original
Faddeev-Jackiw version; the constraints are solved by means of a formal mode
decomposition of the meson $\eta$ and its canonical momentum. In Section 5 we
attack the same problem from the point of view of symplectic quantization 
\cite{wotzasek, montani, barcelos}, where the constraints are not solved but 
incorporated into the symplectic potential; the outcome of this analysis is the
Poisson brackets of the system. The last section contains our conclusions.

%%%%%%%%%%%%%%%%%%%%%%%%%%%%%%%%%%%%%%%%%%%%%%%%%%%%%%%

\section{Preliminaries}
Let $\Phi(x,t)^p$, with $p=1,\ldots,D$ be a set of $D$ classical scalar fields in a $1+d$ dimensional 
manifold $\MM$, and consider the lagrangian density
\beq
\LL=\dot{\HPhi}\,K\,\dot{\Phi}-V(\Phi,\partial_i\Phi;\HPhi,\partial_i\HPhi)
\label{lagden}
\eeq
where dot and $\partial_i$ mean derivative with respect to time $t$ and with
respect to some spatial coordinates $x_i$, $i=1,\ldots,d$ respectively. The 
dagger in $\HPhi$ denotes Hermitian conjugation. The object $K^p_{\,\,\,q}(\Phi)$ is in general a 
hermitian operator that does not contain time derivatives of the fields $\Phi_p$. The potential $V$ 
contains the spatial derivatives of $\Phi$ and $\HPhi$, which we left unspecified (in particular we do not 
require Lorentz invariance). The only condition that $V$ must satisfy is that the classical equations of 
motion have NLW solutions. 

A general NLW solution $\phi(x;\alpha^1,\ldots,\alpha^N)$ will depend on a
number of real parameters $\alpha^a$ which reflect symmetries of the classical equations of motion. In 
a simple case like the kink in $\Phi^4$ theory the translational symmetry of 
the theory leads to static solution of the form $\phi(x-X)$, being $X$ the 
position of the center of mass of the kink. In general the parameters 
$\alpha^a$ are local coordinates on the moduli space ${\mani}$ of static NLWs 
of the theory. The geometry of ${\mani}$ will depend on the potential $V$; we shall not suppose that 
${\mani}$ has any particular structure other than being a smooth manifold with local derivatives
denoted by $\partial_a$. The vectors $\partial_a$ span $T(\mani)$, the 
tangent space of ${\mani}$, at a given point. If $\phi$ is a static NLW,  
$\partial_a\phi$ also satisfies the static equations of motion; in other words, 
the vectors $\partial_a\phi$ are the zero modes of the full time-dependent 
equations of motion.

We shall denote the integration of a function $f$ over a $d$-dimensional 
subspace of $\MM$ by $\int f$, omitting the measure in the integral. This
subspace will be fixed and common to all integrations. If the theory under consideration is defined in 
Minkowski space, this $d$-dimensional subspace will be a maximal space-like submanifold of $\MM$.

The hamiltonian density that corresponds to the original lagrangian density 
(\ref{lagden}) is readily calculated: the canonical momenta are $\Pi=\dot{\HPhi}$ and $\HPi=\dot{\Phi}$, 
and the Hamiltonian 
\beq
\HH=\int\Pi\,\dot{\Phi}+\int\dot{\HPhi}\,\HPi-L=\int \Pi\,K^{-1}\HPi+V.
\label{hamden}
\eeq
The fields $\Phi$ and $\Pi$ are assumed to satisfy the usual Poisson brackets:
\bear
\left\{\Pi(x,t), \Pi(y,t)\right\}&=&\left\{\Phi(x,t), \Phi(y,t)\right\}
=0 \nonumber\\
\left\{\Phi(x,t), \Pi(y,t)\right\}&=&\delta(x-y)
\label{poibra}
\eear

%%%%%%%%%%%%%%%%%%%%%%%%%%%%%%%%%%%%%%%%%%%%%%%%%%%%%%

\section{Collective Coordinates}
Of all the existing procedures to study NLWs we shall be concerned with the
collective coordinates method. In this method \cite{cl, gervais}, the 
semiclassical quantization of a theory that possesses static NLW solutions 
starts with the decomposition of the classical field $\Phi(x,t)$ into two 
parts:
\beq
\Phi(x,t)=\phi[x;\alpha^a(t)]+\eta[x,t;\alpha^a(t)].
\label{total}
\eeq
Generally the field $\phi$ represents a classical solution and $\eta$ the 
quantum fluctuations about it. The choice of a particular classical solution 
$\phi$ will always break some of the symmetries of the theory. For example, if 
the original theory has translation symmetry, the theory built with 
$\phi$ as background will not enjoy translation symmetry. The broken 
symmetries do not altogether disappear from the theory; for each broken
symmetry there will be a collective coordinate in the classical solution 
$\phi$.

The exact meaning of the fields $\phi$ and $\eta$ will depend on what problem we want to solve, and on 
what type of classical solutions we know. If we only have a static solution 
$\phi_0(x;\alpha)$ and we are interested in quantizing the theory in the 
presence of such a classical static field configuration,
then $\phi[x;\alpha^a(t)]=\phi_0(x;\alpha^a)$. In this case the $\alpha^a$ in 
$\phi$ and $\eta$ are time-independent, but we still allow 
for explicit time dependence in $\eta$, typically in the form of oscillatory exponentials 
$e^{i\omega_n t}$ where $\omega_n$ are the normal frecuencies of the system. In other words, $\eta$
will represent small oscillations about the static background $\phi_0$.

If we are interested in quantizing time-dependent field configurations but we
do not know any time-dependent classical solution of the equations of motion
we can still use $\phi_0$. In this situation $\phi[x;\alpha(t)]$ is the same 
function as $\phi_0(x,\alpha)$ but with time-dependent parameters $\alpha^a$. 
It is important to notice that $\phi[x;\alpha(t)]$ is not, in general, a 
time-dependent solution of the classical equations of motion. Its only relation
to $\phi_0(x,\alpha)$ is that we have promoted the parameters $\alpha^a$ to
dynamical, time-dependent variables without changing the functional structure
of $\phi_0$. When we do this we must expect both quantum corrections, and 
kinematical corrections due to $\phi[x;\alpha(t)]$ not being a time-dependent 
classical field configuration. For example, if our theory has Lorentz symmetry
but we take a static solution with time-dependent collective coordinates as our
$\phi[x;\alpha(t)]$ we will find corrections that represent a Lorentz 
contraction of the initial static solution \cite{gervais}. 

Finally, if we have a classical time-dependent solution whose time 
dependence comes from some identifiable collective coordinates $\alpha^a(t)$,
we use it as $\phi[x;\alpha(t)]$. No kinematical corrections should arise in this situation.

Whichever the case, we want $\alpha^a$, with $a=1,\ldots ,N$, and $\eta$ to be 
our new dynamical variables. It is to be noted that we are allowing $\eta$ to depend on the collective
coordinates $\alpha^a$, and assuming at the same time that both $\eta$ and $\alpha^a$ are 
independent dynamical variables. The question of whether the $\alpha^a$ dependence of $\eta$ is redundant or 
not will be shown to be irrelevant in the Faddeev-Jackiw approach. In the symplectic approach the final 
Poisson brackets do detect the dependence of $\eta$ on the collective coordinates; for this reason we shall keep this 
dependence throughout our analysis. 

In the rest of this section we shall formulate the theory in terms of the new variables 
$\alpha$, $\eta$. Inserting the decomposition (\ref{total}) into the lagrangian density (\ref{lagden}) we 
find

\bear
\LL&=&\partial_a(\hphi+\heta)\, \dot{\alpha}^a\,K\,\partial_b(\phi+\eta)\,
\dot{\alpha}^b +\partial_t\heta\,K\,\partial_t\eta \nonumber\\
&+&\partial_a(\hphi+\heta)\,\dot{\alpha}^a\,K\,\partial_t\eta
+\partial_t\heta\,K\,\partial_a(\phi+\eta)\,\dot{\alpha}^a-V,
\label{lagnew}
\eear
where $\partial_t=\partial/\partial_t$. We are interested in the phase space of
this new dynamical system; we must therefore go over to the hamiltonian 
formulation. To this end we define the canonical momenta
\bear
p_a&=&{\partial L\over\partial\dot{\alpha}^a}=\int\left[\partial_a(\hphi+
\heta)\,K\,\partial_b(\phi+\eta)\,\dot{\alpha}^b +\partial_a(\hphi+\heta)\,K\,
\partial_t\eta + \hc \right]\nonumber\\
\pi &=&{\delta L\over\delta\dot{\eta}}=\left[\partial_t\heta+
\partial_a(\hphi+\heta)\,\dot{\alpha}^a\right]\,K.
\label{canmom}
\eear
It is important to observe that the canonical momenta $p_a$ and $\pi$ are not 
independent but related by
\beq
p_a-\int\pi\,\partial_a(\phi+\eta)-\int\partial_a(\hphi+\heta)\,\hpi=0,
\label{cons}
\eeq
In Dirac's terminology the constraint (\ref{cons}) is a first-class constraint that generates the symmetry
\bear
\delta\phi&=&\epsilon^a\,\partial_a\phi,\nonumber\\
\delta\eta&=&-\epsilon^a\,\partial_a\phi,
\label{sym}
\eear
with $\epsilon^a$ an arbitrary $N$-dimensional parameter. This symmetry is
obviously related to the invariance of the decomposition (\ref{total}) under
shifts of $\phi$ and $\eta$ that leave the total field $\Phi$ invariant. This 
indicates that the decomposition (\ref{total}) does not have physical meaning;
in particular, physically meaningful Green functions must be formed with the
total field $\Phi$ \cite{dorey}. 

Besides, the transformation from the initial phase space coordinates $(\Phi, 
\Pi)$ to the new variables $(\alpha^a, p_a, \eta, \pi)$ does not preserve
the canonical 1-form $\omega=\omega_i\,d\xi^i$, (the $\xi^i$ are local 
coordinates in phase space):
\beq
\omega_i\,\dot{\xi}^i=
\int\left( \Pi\,\dot{\Phi}+\hc\right)=p_a\,\dot{\alpha}^a +\int\left(\pi\,
\partial_t\eta +\hc\right)\neq p_a\,\dot{\alpha}^a +\int\left(\pi\,
\dot{\eta}+\hc\right).
\label{cano}
\eeq
The problem is that $\eta$ should appear under a total time derivative (because
$\omega$ is a one-form) instead of under a partial time derivative. Also, the 
non-canonical appearance of $\omega$ implies that the transformation from 
$\Phi$, $\Pi$ to $\alpha^a$, $p_a$, $\eta$, $\pi$ is not a canonical 
transformation\footnote{This problem would not have appeared if $\eta$ had been taken independent
of the collective coordinates $\alpha^a$.}. Besides, the na\"{\i}ve phase space $(p_a, \alpha^a; \pi, 
\eta)$ does not correspond to the true phase space of the theory, because of the 
constraint (\ref{cons})

These facts were first realised by Tomboulis in the particular case of the one-dimensional kink 
\cite{tomboulis}, and by Tomboulis and Woo in their general analysis of soliton quantization
\cite{tw}. These authors applied Dirac's approach to constrained dynamics to 
the system (\ref{lagnew}) subject to the constraint (\ref{cons}). In the next 
sections we shall show that such analysis can be bypassed within the 
Faddeev-Jackiw procedure \cite{fj,jackiw} or its variant \cite{wotzasek,montani,barcelos}. 
At the same time we shall generalise the results of \cite{tw} to include dependence of $\eta$ 
on the collective coordinates $\alpha^a$. Our analysis will also include moduli spaces of NLWs 
with transitive non-abelian groups of transformations, which are relevant to 
monopole quantization \cite{osborn}, especially when the unbroken gauge group is non-abelian
\cite{abouelsaood}.

%%%%%%%%%%%%%%%%%%%%%%%%%%%%%%%%%%%%%%%%%%%%%%%%%%%%%%

\section{Hamiltonian Reduction}
The starting point of this method is a first-order formulation of dynamical systems. The Legendre 
transformation
\beq
L(q,\dot{q})=\omega(q,p)-H(q,p)
\label{legendre}
\eeq
may be understood as a first-order lagrangian in phase space, with the 
hamiltonian playing the role of a potential. We have denoted the symplectic
potential by $\omega$. The object of interest is the symplectic two-form, defined as the exterior
differentiation of the symplectic potential: $\Omega=d\,\omega$. At this point it is convenient to
indicate that the operation $d$ depends on the geometry of phase space. The general situation will now
be described \cite{isham}. Let $\gamma_1, \,\,\gamma_2\in T(\Gamma)$ be two vector fields in 
$\Gamma$. If we denote the natural pairing between a vector $\gamma$ and a 1-form $\omega$ by 
$\langle \omega, \gamma\rangle$, and the acting of a two-form $\Omega$ on two vectors by
$\Omega(\gamma_1,\gamma_2)$, we define the operation $d$ acting on $\omega$ as
\beq
d\omega(\gamma_1,\gamma_2)=\gamma_1 \langle\omega,\gamma_2\rangle-
\gamma_2 \langle\omega,\gamma_1\rangle-\langle \omega, [\gamma_1,\gamma_2]\rangle.
\label{true}
\eeq
Particularly interesting examples of dynamical systems are those with a configuration space $Q$ 
isomorphic to a connected Lie group $G$. For these systems the vector fields $\gamma$ restricted to 
the tangent space of $Q$ can always be taken as the ``push forward" of the Lie algebra of $G$. In these
cases the symplectic form can be written in components as
\beq
\Omega_{ij}={\partial\over\partial\xi^i}\,\omega_j-
{\partial\over\partial\xi^j}\,\omega_i -F_{ij}^{\,\,\,k}\omega_k
\label{symplectic}
\eeq
where we have used the structure constants of the algebra of vector fields:
\beq
[\gamma_i,\gamma_j]=F_{ij}^{\,\,\,k}\gamma_k.
\label{lie}
\eeq
If the configuration space admits a transitive abelian group of transformations, the ``non-abelian" term in 
eq.(\ref{symplectic}) vanishes. Also, this term may not be present if the phase space coordinates are chosen not to
include the vectors $\gamma$. This point will be ellaborated below for a configuration space that is a group
manifold.

If the symplectic matrix $\Omega_{ij}$ is non-singular its inverse, denoted by
$\Omega^{ij}$, will exist. In this situation there are no true constraints on 
the system, the Poisson brackets of two functions $F(\xi)$, $G(\xi)$ are given by
\beq
\left\{F(\xi), G(\xi)\right\}=\partial_iF\,\Omega^{ij}\,\partial_jG
\label{pobr}
\eeq
and we have a complete hamiltonian description of the dynamics. When the 
symplectic matrix is singular the system is constrained. This is due 
to the existence of zero modes of the symplectic matrix. If we denote these
zero modes by $Z^n$, being $n$ an index that enumerates the different zero
modes, the hamiltonian equations of motion imply that
\beq
(Z^n)^i\,{\partial\over\partial\xi^i}\,H=0.
\label{eqsofm}
\eeq
These equations involve no time derivatives, and therefore correspond to
the constraints of the problem. In the Faddeev-Jackiw approach 
\cite{fj,jackiw} these constraints are simplified by means of a 
Darboux transformation from the original variables $\xi^i$ to new ones 
$p_k, q^l, z_m$. In the new variables the symplectic potential 
$\omega$ takes the form $p_i\,d{q}^i$, while the hamiltonian may 
also depend on the $z_m$. Equation (\ref{eqsofm}) will now correspond to the 
equations of motion for the variables $z_m$ 
\beq
{\partial\over\partial z_m}H(p,q,z)=0,
\label{coco}
\eeq
which can be used to evaluate the $z$'s in terms of the $p$'s and $q$'s 
unless $H$ depends linearly on some $z$'s. If this elimination is performed 
one is left with an expression linear in the surviving $z$ variables, 
\beq
L=p_i\dot{q}^i-H(p,q)-z_m h^m(p,q)
\label{haha}
\eeq
and the only true constraints are $h^m = 0$. These constraints must 
be solved and the solution used to reduce the number of degrees of freedom in 
the lagrangian. If the new, reduced symplectic matrix is still singular the 
procedure starts again, until one finally arrives at an unconstrained 
lagrangian. We wish to point out that sometimes the system is given directly 
in the form (\ref{haha}). This happens, for example, in electrodynamics; Gauss' Law appears as a 
constraint in the lagrangian from the beginning. 

After this brief review of the Faddeev-Jackiw treatment of constrained systems
we turn back to our NLWs. We shall first consider the problem of the non
conservation of the symplectic potential. We can write $\omega$ as follows:
\bear
\omega_i\,\dot{\xi}^i&=&p_a\,\dot{\alpha}^a+\int\left(\pi\,\partial_t\eta+\hc\right)
\nonumber\\
&=&\left[p_a-\int\left(\pi\,\partial_a\eta+\hc\right)\right]\,\dot{\alpha}^a + \int\left(\pi\,\dot{\eta}+\hc\right)\\
&=& \tilde{p}_a\,\dot{\alpha}^a +\Re\int\pi\,\dot{\eta}, \nonumber
\label{step}
\eear
where we have defined a new momentum $\tilde{p}$ as
\beq
\tilde{p}_a=p_a-\int\left(\pi\,\partial_a\eta+\hc\right).
\label{tilde}
\eeq
In the new variables $\,\tilde{p}_a$, $\alpha^a$, $\pi$ and $\eta$, the 
symplectic potential $\omega$ is in canonical form. Therefore we should adopt 
these new variables as local coordinates in the phase space of our 
system. This is, of course, only an intermediate step, since we still have to deal with 
the constraint (\ref{cons}). 
\subsection{Solving the Constraints}
Let us write the lagrangian in first-order form:
\bear
L&=&\tilde{p}_a\,\dot{\alpha}^a+\int\left(\pi\,\dot{\eta}+\hc\right)-H \nonumber\\
H&=&\int\pi\,K^{-1}\,\hpi + \int V.
\label{lagdosa} \nonumber
\eear
The system (\ref{lagdosa}) together with the constraint (\ref{cons}) is already
in the form (\ref{haha}), so no Darboux transformation is needed to expose the 
constraint. The task is therefore to solve the constraint (\ref{cons}). The obvious way to proceed 
is to eliminate $p_a$ in favour of $\pi$ and the $\alpha$'s, 
\beq
p_a=\int\left(\pi\partial_a(\phi+\eta)+\hc\right),
\label{solve}
\eeq
but this leads to a canonical set of variables that undoes the decomposition
(\ref{total}), \ie\ requires $\Phi$ and $\Pi$ as canonical variables. The reason for this is, roughly, that 
in the solution (\ref{solve}) $\phi$ and $\eta$ play a symmetric role, so that there
is no way to separate them in the resulting lagrangian. The alternative is to 
eliminate $\pi$ in terms of $p_a$ and $\alpha^a$. It will be convenient to introduce the notation
\bear
\mu_{ab}&=&\int\partial_a\hphi\,K\,\partial_b\phi, \nonumber\\
\xi_{ab}&=&\int\partial_a\hphi\,K\,\partial_b\eta. 
\label{muab}
\eear
Let us decompose the momentum $\pi$ into ``transverse'' and 
``longitudinal'' components with respect to the directions $\partial_a\phi$:
\beq
\pi=\tilde{\pi}+\half\left[p_a-\int\left(\tilde{\pi}\,\partial_a\eta+\hc\right)\right]
\left(\mu+\xi\right)^{ab}\partial_b\hphi\,K,
\label{tildepi}
\eeq
where $\tilde{\pi}$ is transverse in the sense that
\beq
\psi_a\equiv\int\tilde{\pi}\partial_a\phi=0.
\label{trans}
\eeq
The constraint is now $\psi_a=0$. It is also important to notice that the 
transversality of $\tilde{\pi}$ is relative to $N$ functions $\partial_a\phi$. 
This contrasts with the situation in electrodynamics, where the electric field 
${\bf E}$ must also be decomposed in transversal and longitudinal components
\beq
{\bf E}={\bf E}_T+{\bf E}_L,
\label{electro}
\eeq
but the transversality of ${\bf E}_T$ is defined as $\nabla\cdot{\bf E}_T=0$ 
without involving other fields. 

Now we must solve the constraint (\ref{trans}). A way to proceed is to choose a base ${\cal B}$ of 
orthonormal functions in the Hilbert space $\hi$ that should contain $N$ vectors $f_a$ that diagonalize $\mu_{ab}$. 
These $N$ vectors must exist because $\mu_{ab}$ is defined hermitian. Let us write the elements of this base as
\bear
{\cal B}&=&\{ f_a; f_i\} \quad\quad a=1,\ldots,N\quad\quad 
i=N+1,\ldots,\infty \nonumber\\
(f_m,f_n)&\equiv&\int f_m^{\dag}\,K\,f_n=\delta_{mn}\quad\quad 
m,n=1,\ldots,\infty
\label{base}
\eear
where $(\cdot,\cdot)$ is the inner product defined in $\hi$. We  can always consider that the $f_a$ are proportional to 
$\partial_a\phi$. Having introduced the basis ${\cal B}$, we solve the constraint (\ref{trans}) by restricting 
$\tilde{\pi}$ to be in the subspace of $\hi$ that is orthogonal to the $f_a$,
\beq
\tilde{\pi}(x,t)=\sum_{n=N+1}^{\infty}\,c_n(t)\,f_n^{\dag}\,K,
\label{deco}
\eeq
where the $f_i$ will depend on the collective coordinates $\alpha^a(t)$ while 
the coefficients $c_n$ will depend only on time. Using the decomposition (\ref{tildepi}) 
and the mode decomposition (\ref{deco}) we write the
symplectic potential, up to a total time derivative, as
\beq
\omega_i\,\dot{\xi}^i =\hat{p}_a\,\dot{\alpha}_a + 
\sum_{n=N+1}^{\infty}\int \left(c_n f_n^{\dag}\,K\,\dot{\eta}+\hc\right)-\dot{v}^a\,\chi_a-\chi^{a\,\,\dag}\,
\dot{v}_a^{\dag},
\label{coms}
\eeq
where the following definitions have been used:
\bear
\hat{p}_a&=&\tilde{p}_a-\left\{\half\left[p_b-\int\left(\tilde{\pi}\,\partial_b\eta+\hc\right)\right]
(\mu+\xi)^{bc}\int\partial_a\partial_c\hphi\,K\,\eta+\hc\right\},\nonumber\\
\chi_a&=&\int\partial_a\hphi\,K\,\eta,\\
v_a&=&\half\left[p_b-\int\left(\tilde{\pi}\,\partial_b\eta+\hc\right)\right](\mu+\xi)^{ba}.
\nonumber
\label{fede}
\eear

We can also decompose the field $\eta$ in the base ${\cal B}$. At the same 
time we observe that the symplectic potential is not in canonical form due to
the last two terms in (\ref{coms}). This can be solved by assuming $\chi_a=0$. In
Dirac's terminology this is a second-class constraint that corresponds to
a ``gauge condition''. In this terminology, we are choosing a gauge where  
$\eta$ is orthogonal to the vectors $\partial_a\phi$:
\beq
\eta(x,t)=\sum_{n=N+1}^{\infty}\,q_n(t)\,f_n(x). 
\label{decoeta}
\eeq
where the explicit time dependence lies in the coefficients $q_n$ and the 
dependence on the collective coordinates is in the functions $f_n$, as in the
decomposition of $\tilde{\pi}$. This choice is sometimes called ``rigid gauge''
\cite{dorey}. Using the orthonormality of the elements of the
base ${\cal B}$, the symplectic potential is now
\beq
\omega_i\dot{\xi}^i=\hat{p}_a\,\dot{\alpha}^a+\sum_{n=N+1}^{\infty}\left(c_n\,
\dot{q}_n +\hc\right)+\int\left(\tilde{\pi}\,\partial_a\eta+\hc\right)\,\dot{\alpha}_a.
\label{final1}
\eeq
It is now clear that a further redefinition of the momentum will render the
symplectic potential in Darboux form. The new momentum is
\beq
\bar{p}_a=\hat{p}_a+\int\left(\tilde{\pi}\,\partial_a\eta+\hc\right),
\label{momentum}
\eeq
which leads to the final form of the symplectic potential:
\beq
\omega_i\,\dot{\xi}^i=\bar{p}_a\,\dot{\alpha}^a+\sum_{n=N+1}^{\infty}\left(c_n\,\dot{q}_n+\hc\right). 
\label{sympl}
\eeq
We wish to remark here that the orthonormality of the elements of the base ${\cal B}$ makes
irrelevant the dependence of $\eta$ on the collective coordinates. It is the normal modes $c_n$, $q_n$ that act 
as coordinates and momenta, and these do not depend on the collective coordinates. 

\subsection{Gauge Invariance}
The decomposition (\ref{decoeta}), that corresponds to $\chi_a=0$, is not the most general way to put 
the symplectic potential $\omega$ in canonical form. Here we shall show that there is a large class of 
``gauges'' beyond the rigid one that lead to the same $\omega$.  

In the discussion leading to (\ref{sympl}) our choice was to take $\eta$ orthogonal to the zero nodes 
$\partial_a\phi$. This choice was motivated by the decomposition of $\pi$ into
transversal and longitudinal components with respect to the $\partial_a\phi$.
However, as shown in \cite{dorey}, the orthogonality of $\eta$ can de defined 
with respect to any set of linearly independent vectors $g_a\in T(\mani)$ with 
$a=1,\ldots,N$. In order to see this possibility we must solve again the 
constraint (\ref{cons}) by means of a more general decomposition of $\pi$.

The vectors $g_a$ can be used to complete a basis ${\cal B}'=\{g_a;g_i\}$ 
with $i>N$ and $(g_a,g_i)=0$. This new basis is not simply a rotation of the
original basis ${\cal B}$; $g_a$ and $\partial_a\phi$ may have different
functional form. Let us define now the matrix
\beq
\Lambda_{ab}=\int g_a^{\dag}\,K\,\partial_b(\phi+\eta).
\label{lambda}
\eeq
We need this matrix to be non-singular. Then we denote its inverse by 
$\Lambda^{ab}$ and write the solution of the constraint (\ref{cons}) as
\bear
\pi&=&\tilde{\pi}'+\half\left[p_a-\int\left(\tilde{\pi}\,\partial_a(\phi+\eta)\right)+\hc\right]
\Lambda^{ab}g_b^{\dag}, \nonumber\\
\tilde{\pi}'&=&\sum_{i=N+1}^{\infty}a_i(t)\,g_i^{\dag}\,K.
\label{solu2}
\eear
The new symplectic potential will be in Darboux form if $\eta$ is constrained
to be orthogonal to the $g_a$. This is achieved by means of a new mode 
decomposition of $\eta$ which should exclude the $g_a$. Thus we have verified 
that it is possible to define transversality with respect to a general set
of $N$ vectors $g_a\in T(\mani)$ and preserve the canonical structure of the
theory at the same time.

An even more general choice would be to allow for a constant longitudinal
component of $\eta$, not necessarily zero. In our language this corresponds
to the weaker constraint 
\beq
\chi_a=K(x)\quad\quad {\rm with} \quad\quad \dot{K}(x)=0.
\label{weak}
\eeq
If this is possible, the symplectic potential will be in Darboux form 
up to total time derivatives, which can be dropped. In terms of a mode 
decomposition, (\ref{weak}) amounts to the following condition on the 
longitudinal modes of the meson field $\eta$:
\beq
{d\over dt}\left[ q_a(t)\, \|\partial_a\phi\|^a\right] =0\quad\quad
{\rm or}\quad\quad{\dot{q}_a\over q_a}=-{2 \over \|\partial_a\phi\|}
{d\over dt}\|\partial_a\phi\|.
\label{condition}
\eeq
If we insist that the coefficients $q_n(t)$ of the mode decomposition depend 
explicitly on $t$ and not on $\alpha^a$ we must conclude that the norms
$\|\partial_a\phi\|$ must be independent of the collective coordinates 
$\alpha^a$. In the more general basis ${\cal B}'$ the condition is that the norms $\|g_a\|$, 
$a=1,\ldots,N$ should be independent of time. If this cannot be satisfied in a particular model, we must 
take the restricted condition (\ref{decoeta}). Many models of interest, however, do 
satisfy this extra condition. If the collective coordinates correspond to the 
center of mass of a soliton, the dependence on the $\alpha^a$ will be of the 
form $\phi(x^a-\alpha^a)$ and therefore the integration over the coordinates 
$x^a$ will eliminate the dependence on $\alpha^a$ in $\|\partial_a\phi\|$. 
This also happens if $\mani$ is an abelian group manifold; the
collective coordinates appear as parameters in the exponential map of the
group and cancel in $\|\partial_a\phi\|$. When the weak condition (\ref{weak}) 
can be taken the final symplectic potential $\omega$ will be equivalent to the 
symplectic potential arising from the strong constraint $\chi_a=0$, since the 
difference amounts to a total time derivative.

\subsection{Poisson Brackets}
After eliminating the unphysical degrees of freedom introduced by the decomposition (\ref{total}), the 
phase space is spanned by $(\bar{p}_a,\alpha^a, c_n, d_n, c_m^{\ast}, d_m^{\ast})$. The 
symplectic matrix $\Omega$ is easily found to be
\bear
\Omega_{ij}=\left(
\begin{tabular}{cccc}0&$I_{ac}$&0&0\\
$-I_{db}$&0&0&0\\
0&0&0&$\I_{mn}$\\
0&0&$-\I_{pq}$&0
\end{tabular}
\right).
\label{novaf1}
\eear
The notations $I$ and $\I$ stand, respectively, for the $N\times N$ unit matrix
and for the $\aleph_0\times\aleph_0$ unit matrix. The inverse of $\Omega$ 
exists and determines the fundamental Poisson brackets between phase space 
coordinates:
\bear
\{\alpha^a, \alpha^b\}&=&\{\bar{p}_a, \bar{p}_b\}=0\nonumber\\
\{\alpha^a, \bar{p}_b\}&=&\delta_b^{\,\,a},\nonumber\\
\{q_n, c_m\}&=&\I_{mn},\\
\{q_n, q_m\}&=&\{c_n, c_m\}=0 .\nonumber\\
\{\bar{p}_a, c_n\}&=&\{\bar{p}_a, q_n\}=0
\label{poi}
\eear
Canonical Quantisation of this reduced dynamical system would proceed along 
the usual lines. Thus the Faddeev-Jackiw method applied to the quantization of 
non-linear waves reproduces the usual decomposition in longitudinal and 
transverse modes with respect to the background configuration 
$\phi[x;\alpha(t)]$ that appears in \cite{gj,tw}. 

\subsection{The Case of a Group Manifold}
In the Poisson brackets above, the momenta $\bar{p}_a$ commute with themselves because the momenta $\bar{p}_a$ 
are not the vector fields on the group manifold introduced in (\ref{true}). In order to show the relation between the 
momenta $\bar{p}_a$ and the non-commuting vector fields on a group manifold (see the Appendices of \cite{jackiw2} and 
\cite{wadia}) let us parametrize the field $\Phi$ and the matrix $K$ in (\ref{lagden}) as
\bear
\Phi(x,g)&=&h(x)\,g,\quad\quad g\in G,\nonumber\\
K(\Phi)&=&k(x)\,K_g
\label{par}
\eear
where $G$ is a non-abelian Lie group and $K_g$ acts on $G$ only. Up to constants, the part of the lagrangian that is relevant 
to the dynamics in $G$ is
\beq
L_G=\half\tr \left[ \dot{g}^{-1}\,K_g\,\dot{g}\right]=-\half\tr\left[ \dot{g}\,g^{-1}\,K_g\,\dot{g}\,g^{-1}\right].
\label{lagg}
\eeq
If the group element $g$ depends on time through some collective coordinates $\alpha^a$ its time dependence can
be written as
\beq
\dot{g}=\partial_a\,g\,\dot{\alpha}^a. 
\label{gtime}
\eeq
We can now define two momenta associated to the group element $g$: the ``intrinsic" momentum $J_a$ and the
``canonical" momentum $p_a$
\bear
J_a&=&\tr\left[{\delta L\over \delta(\dot{g}\,g^{-1})}\,T_a\right]=\tr\left[T_a\,K_g\,\dot{g}\,g^{-1}+
\dot{g}\,g^{-1}\,K_g\,T_a\right], \nonumber\\
p_a&=&{\delta L\over \delta\dot{\alpha}^a}=\tr\left[ \partial_a g\,g^{-1}\,K_g\,\dot{g}\,g^{-1}+
\dot{g}\,g^{-1}\,K_g\,\partial_a g\,g^{-1}\right].
\label{intrinsic}
\eear
These two momenta are related through the ``vierbein" $E_a^{\,\,b}$ on the group manifold, which we define as
\beq
i\,T_a=E_a^{\,\,b}\,\partial_bg\,g^{-1}.
\label{vierbein}
\eeq
This object satisfies the integrability condition of the Lie group:
\beq
E_a^{\,\,c}\,\partial_c E_b^{\,\,d}-E_b^{\,\,c}\,\partial_c E_a^{\,\,d}=f_{ab}^{\,\,\,c}\,E_c^{\,\,d}.
\label{inte}
\eeq
The relation between the intrinsic and the canonical momentum is $J_a=-i\,E_a^{\,\,b}\,p_b$. It is now easy to prove that
$J_a$ satisfy the Poisson brackets thet we expect from vectors on a non-abelian group manifold:
\beq
\left\{ J_a, J_b\right\}=i\,f_{ab}^{\,\,\,c}\,J_c.
\label{jota}
\eeq

%%%%%%%%%%%%%%%%%%%%%%%%%%%%%%%%%%%%%%%%%%%%%%%%%%%%%%%
%%%%%%%%%%%%

\section{Symplectic analysis}
Let us reconsider the general constraint equation (\ref{eqsofm}). In order to 
solve these constraints we must be able to construct the Darboux transformation
to canonical coordinates. If that is possible, we still have to solve 
the new constraints (\ref{coco}), which means that some variables 
have to be written in terms of a reduced set of coordinates. 

When that direct approach is not feasible we can resort to Dirac's 
procedure, or adopt the method proposed in \cite{wotzasek, montani}. The main 
idea in this method is to include the ``unsolvable'' constraints, that we
shall call $C_k$, into the lagrangian by means of Lagrange multipliers 
{\sl that are velocities\/}. This has two effects on the theory:
\begin{enumerate}
\item The constraints are now part of a new, enlarged symplectic potential. 
We are then introducing a new symplectic matrix that, if all constraints are
taken into account, will be regular. 
\item The Poisson brackets defined in this enlarged phase space are Dirac 
brackets \cite{wotzasek, montani, barcelos}.
\item We are shifting the constraints to the tangent space of the phase space 
of the dynamical system. In other words, the lagrangian equations of motion
of the Lagrange multipliers ensure the stability of the constraints under
time evolution:
\beq
\dot{C}_k=0.
\label{omega}
\eeq
\end{enumerate}
We should also impose the initial condition $C|_{t=0}=0$ in order to make
equivalent the dynamics of the new system to that of the system before 
introducing velocity Lagrange multipliers. It is to be 
noted, however, that in the original Faddeev-Jackiw method one is solving
the constraints without requiring that they should be preserved under time 
evolution. In all known cases it is not necessary to introduce that stability 
condition by hand; the dynamical system, if consistent, preserves automatically
the constraints under time evolution \cite{private}. The question of under what
conditions that is true deserves further examination but will not be 
addressed here. 

We shall, in this section, apply this idea of enlarging (rather than reducing) 
the symplectic potential with velocity Lagrange multipliers to the collective 
coordinates approach to NLWs. In the previous section the constraint 
(\ref{trans}) was solved by choosing 
a particular basis ${\cal B}$ of functions. If we do not want to choose any
basis, or if we do not know how to define it in a specific situation, we can
still apply the modified procedure just described to quantize NLWs. The 
new, enlarged symplectic potential, denoted by $\omega'$, is
\beq
\omega'=\hat{p}_a\,\dot{\alpha}_a+\left[\int\tilde{\pi}\,\dot{\eta}+
\chi_a\,\dot{\lambda}^a+\psi_a\,\dot{\sigma^a} +\hc \right],
\label{lagfinal}
\eeq
where in the Lagrange multiplier $\dot{\lambda}_a$ has absorbed the 
$-\dot{v}_a$ present in (\ref{coms}). The new variables $\sigma$ and 
$\lambda$ have transferred the constraints to the symplectic potential. We 
can now verify that the new extended symplectic matrix is regular, and then 
define Poisson brackets between the coordinates of the new phase space without 
having to resort to Dirac's method. The new symplectic matrix is
\bear                                                                                            
\Omega_{ij}=\left(
\begin{tabular}{cccccc|cccc}0&$I$&0&0&0&0&0&0&0&0\\
$-I$&0&0&0&0&0&$\Xi^T$&$\Delta$&$\Xi^{\dag}$&$\Delta^*$\\
0&0&0&$\delta$&0&0&0&$\partial\phi$&0&0\\
0&0&$-\delta$&0&0&0&$\partial\phi^{\dag}K$&0&0&0\\
0&0&0&0&0&$\delta$&0&0&0&$\partial\phi^{\dag}$\\
0&0&0&0&$-\delta$&0&0&0&$K\partial\phi$&0\\
\hline
0&$-\Xi$&0&$-\partial\phi^{\dag}K$&0&0&0&0&0&0\\
0&$-\Delta^T$&$-\partial\phi$&0&0&0&0&0&0&0\\
0&$-\Xi^*$&0&0&0&$-K\partial\phi$&0&0&0&0\\
0&$-\Delta^{\dag}$&0&0&$-\partial\phi^*$&0&0&0&0&0
\end{tabular}
\right).
\label{novaf}
\eear
where we have eliminated subindexes (as in $\partial\eta$) in order to 
simplify the notation, and $\Delta$ and $\Xi$ are $N\times N$ matrices with elements
\bear
\Delta_{ab}&=&\int\left( \tilde{\pi}\,\partial_{ab}\phi+\partial_a\tilde{\pi}
\,\partial_b\phi\right), \nonumber\\
\Xi_{ab}&=&\xi_{ab}+\int\partial_b\partial_a\hphi\,K\,\eta.
\label{delta}
\eear
As announced, the extended $\Omega_{ij}$ is non-singular. The lines separate the 
contributions from the physical variables $\tilde{p}$, $\alpha$, $\tilde{\pi}$ 
and $\eta$ from the contributions due to the constraints and Lagrange 
multipliers. The inverse of $\Omega_{ij}$ provides the basic Poisson brackets 
between the canonical variables of $L'$,
\bear
\{\hat{p}_a,\hat{p}_b\}_\dir &=&\left[\Delta\mu^{-1\,\,T}
\Xi-\Xi^T\mu^{-1}\Delta^T+\Delta^*\mu^{-1\,\,T}\Xi^*-\Xi^{\dag}\mu^{-1}\Delta^{\dag}
\right]_{ab}\equiv M_{ab},\nonumber\\
\{\alpha^b, \hat{p}_a\}_\dir&=&\delta_a^{\,\,b},\nonumber\\
\{\hat{p}_a, \tilde{\pi}\}_\dir&=&-\partial_a\tilde{\pi}
+\Delta_{ab}\mu^{cb}\partial_c\phi^{\dag}K,\nonumber\\
\{\hat{p}_a,\eta\}_\dir&=&-\partial_a\eta+\Xi_{ba}\mu^{cb}\partial_c\phi,\nonumber\\
\{\hat{p}_a,\sigma^b\}_\dir&=&-\Xi_{ca}\mu^{cb},\nonumber\\
\{\hat{p}_a,\lambda^b\}_\dir&=&\Delta_{ac}\mu^{cb},\nonumber\\
\{\tilde{\pi}(x),\eta(y)\}_\dir&=&-\delta(x-y)+\partial_a\phi(x)^{\dag}\,K\,
\mu^{ab}\,\partial_b\phi(y)\nonumber\\
\{\tilde{\pi}(x),\sigma_a\}_\dir&=&-\partial_b\phi(x)^{\dag}\,K\,\mu^{ba},\nonumber\\
\{\eta(x),\lambda^a\}_\dir&=&-\mu^{ab}\partial_b\phi,\nonumber\\
\{\lambda^a,\sigma^b\}_\dir&=&-\mu^{ab}\,\, .
\label{poitw}
\eear
The subindex $D$ indicates that these are Dirac brackets. The rest of the Poisson brackets are
zero or can be obtained by hermitian conjugation. It is easy to check that the constraints $\xi_a$
and $\psi_a$ have vanishing brackets with all the variables, which ensures that the constraints
do not evolve in time. 

If the configuration space of the theory is a non-abelian group manifold, the intrinsic momenta $J_a$ defined in the 
previous section satisfy now a more complicated Lie algebra:
\beq
\{ J_a, J_b\}=i\,f_{ab}^{\,\,\,c}\,J_c-E_a^{\,\,c}\,M_{cd}\,E_b^{\,\,d}.
\label{newalg}
\eeq

%%%%%%%%%%%%%%%%%%%%%%%%%%%%%%%%%%%%%%%%%%%%%%%%%%%%%%%
%%%%%%%%%%%%%%%%%%%%

\section{Conclusions}
It was the purpose of this article to apply the modern approach to constrained
systems pioneered by Faddeev and Jackiw \cite{fj,jackiw} to non-linear waves. The alternative method 
due to Wotzasek, Montani and Barcelos-Neto \cite{wotzasek, montani, barcelos} has also been applied
to the same problem.

The original Faddeev-Jackiw method leads to a reduced phase space with only the
``true'' degrees of freedom. Moreover, after a Darboux transformation the
symplectic form, and therefore the Poisson brackets, will be canonical. Use
of this approach leads to the necessity of introducing a formal mode 
decomposition of the meson $\eta$ and the canonical momentum $\tilde{\pi}$. 
The final structure of the dynamical system is a generalisation of the Christ-Lee
version of the collective coordinate formalism \cite{cl}. 

In the modified procedure we do not solve the constraints; these are absorbed
into the symplectic potential by means of velocity Lagrange multipliers. The
total phase space must now include the Lagrange multipliers. The new, expanded
symplectic structure will be regular if all constraints are taken into account.
The inverse of the symplectic matrix determines the Poisson (Dirac) structure 
of the phase space. When applied to the collective coordinate analysis of NLWs,
we find results that reduce to those of \cite{tomboulis, tw} when the meson
$\eta$ does no depend on the collective coordinates.

From a practical point of view it is clear that the simplest approach is the
original Faddeev-Jackiw Hamiltonian Reduction in the sense that it leads to fewer degrees
of freedom, which furthermore are the physical ones. Its only disadvantage is
that it relies on a mode decomposition of the meson field and its canonical
momentum. This approach will therefore loose manifest covariance in field 
space. It may also happen that the elimination of the unphysical 
longitudinal degrees of freedom obscures some symmetries of the system. 
The symplectic approach is much less suitable for practical calculations
but does not require a particular choice of basis to decompose the
meson field. It is therefore more useful when we want to investigate general
properties of the dynamics of small fluctuations around NLWs, like symmetries
and global aspects of the phase space.

\vskip 2cm
\begin{center}
{\bf Acknowledgements}
\end{center}
I am indebted to Professor Roman Jackiw for his advice and interest in this 
work, and to Professor John Negele and the Center for Theoretical Physics, 
where part of this work was done, for hospitality. I would like to thank the 
Spanish Ministerio de Educaci\'on y Ciencia and the UK Engeneering and Physical
Sciences Research Council for financial support.

%%%%%%%%%%%%%%%%%%%%%%%%%%%%%%%%%%%%%%%%%%%%%%%%%%%%%%%
%%%%%%%%%%%%%%%%%%%%%%%

\end{document}